\documentclass[12pt]{article}
\def\iue{$IUE$}
\def\units{$10^{-10}\mathrm{erg}\hspace{0.8mm}\mathrm{s}^{-1}\mathrm{cm}^{-2}\mathrm{\AA}^{-1}$}
\def\mymail{n\_\hspace{0.1cm}sokolov58@mail.ru}
\usepackage{natbib}
\usepackage{graphicx}
\newcommand\sect[1]{\bigskip{\bf\noindent #1}\medskip}
\begin{document}
\pagestyle{plain}
\setcounter{page}{1}
\begin{center}\large\bf
%-----------------------------------------------------------------
%  Title:
%-----------------------------------------------------------------
Variability of C~{\sc III} and Si~{\sc III} lines in the ultraviolet
spectral region \\
of the magnetic Bp star a Centauri
\end{center}
\begin{center}\sc
%-----------------------------------------------------------------
%  Authors:
%  e.g. A. Author$^1$, B. Author$^2$, and C. Author$^1$
%-----------------------------------------------------------------
Sokolov N.A.
\end{center}
\begin{center}
%-----------------------------------------------------------------
%  Institute:
%  e.g. $^1$ Institute of A, Japan <E-mail address A>\\
%       $^2$ Institute of B, Japan <E-mail address B>
%-----------------------------------------------------------------
 Central Astronomical Observatory at Pulkovo,
 St. Petersburg 196140, Russia \\
 E-mail: {\mymail} \\
\end{center}
%-----------------------------------------------------------------
%  Main Text
%-----------------------------------------------------------------
\abstract{
The variability of twice ionized lines of carbon and silicon in
the ultraviolet spectral region of the magnetic Bp star $a$~Centauri
is investigated. This study is based on the archival {\it International
Ultraviolet Explorer\/} data obtained through the large aperture and
in the low-dispersion mode. A comparison of the average {\iue}
high-dispersion spectrum of $a$~Cen with full synthetic spectrum as well
as those including only lines of one element showed that six C~{\sc iii}
and six Si~{\sc iii} lines are responsible for the depressions of the flux
at $\lambda\lambda$\,1175.5 and 1300\,\AA, respectively.
Investigation of the variability of flux in the core of depression at $\lambda$\,1775.5\,\AA\  indicate that the fluxes do not vary within errors
of measurements. On the other hand, the fluxes in the core of depression
at $\lambda$\,1300\,\AA\ varies significantly with amplitude of $\sim$0.2~mag. Moreover, the variability of this depression are in anti-phase with helium
lines in the visual spectral region.
}
%-----------------------------------------------------------------

\sect{1 Introduction}

$a$~Centauri (HD~125823, HR~5378, V761 Cen) belongs to the Si-subclass. The star is a striking helium variable, ranging in helium spectral type from B2 to B8 with a period of 8.81 d \citep{Jaschek_et_al_1968}. \citet{Norris_1968} found that
the He~{\sc i}~$\lambda$\,4471 varies from 300 to 2200 m\AA\ and the changes are so conspicuous that $a$~Cen can be considered a He-weak star at one extremum He-rich at the other one. In detailed study of this star \citet{Norris_1971} showed the equivalent widths of all neutral helium lines have similar extreme variations. On the other hand, the metallic lines show a very little or no changes of their intensity. In particular, the lines of Si~{\sc ii} show no variation, while the lines of Si~{\sc iii} show the variation in anti-phase with respect to the helium lines.

Our goal is to investigate the variability of C~{\sc iii} and Si~{\sc iii} lines in the ultraviolet spectral region using low-dispersion spectra which are presented in the {\iue} Newly Extracted Spectra (INES) database from the {\iue} satellite.

\sect{2 Observational data and analysis}

The series of {\iue} spectra of $a$~Cen obtained with Short Wavelength Prime (SWP) camera was received from the INES archive.
In all cases, the spectra were obtained through the large aperture (9.5"~x~22") and in the low-dispersion mode ($\sim$~6\,\AA). All SWP spectra were obtained with exposure time 2.148 second. Finally, 13 SWP spectra were investigated and distributed quite smoothly over the period of rotation.
Additionally, three spectra (SWP~14071, SWP~14080, SWP~14088) obtained in May 1981 were received from the INES archive.
These spectra were obtained through the large aperture in the high-dispersion mode. In our study we used them only for identification of the observed depressions in
the spectrum of $a$~Cen.

To analyze the {\iue} spectra of $a$~Cen the linearized least-squares method was used. An attempt was made to describe the light curves
in a quantitative way by adjusting a Fourier series.
Based on the fact that the light variations of $a$~Cen are well approximate by single-wave \citep{Catalano_1996}, we used for each light curve a least square
fit with a function of the type:
\begin{equation}
F(\lambda, T^{'})=A_{0}(\lambda) +
A_{1}(\lambda)\cos(\omega~T^{'} +\phi_{i}(\lambda)),
\label{equation_2}
\end{equation}
where $F(\lambda, T^{'})$ is a flux for the given $\lambda$, $T^{'}$=~$T$~-~$T_{0}$
and $\omega$~=~2$\pi$/$P$. The $T_{0}$ and $P$ are zero epoch and rotational
period of the ephemeris, respectively. The coefficients $A_{0}(\lambda)$ of the fitted curves define the average distribution of energy over the cycle of the variability while the coefficients $A_{1}(\lambda)$ define the semi-amplitude of
the flux variations for the given $\lambda$.

\begin{figure*} [t]
\centerline{\includegraphics[width=120mm, angle=0]{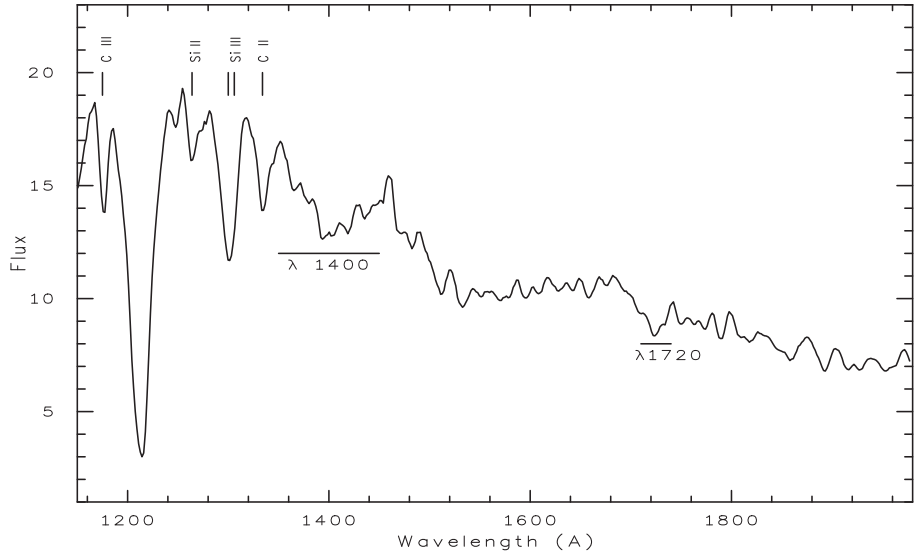}}
\caption{The average distribution of energy in {\units} for $a$~Cen.
 The conspicuous depressions and features are shown by vertical and
 horizontal lines, respectively.}
\label{average}
\end{figure*}
\begin{figure}[t]
%\vspace{-4.5cm}
\centering
\resizebox{0.49\hsize}{!}{\includegraphics{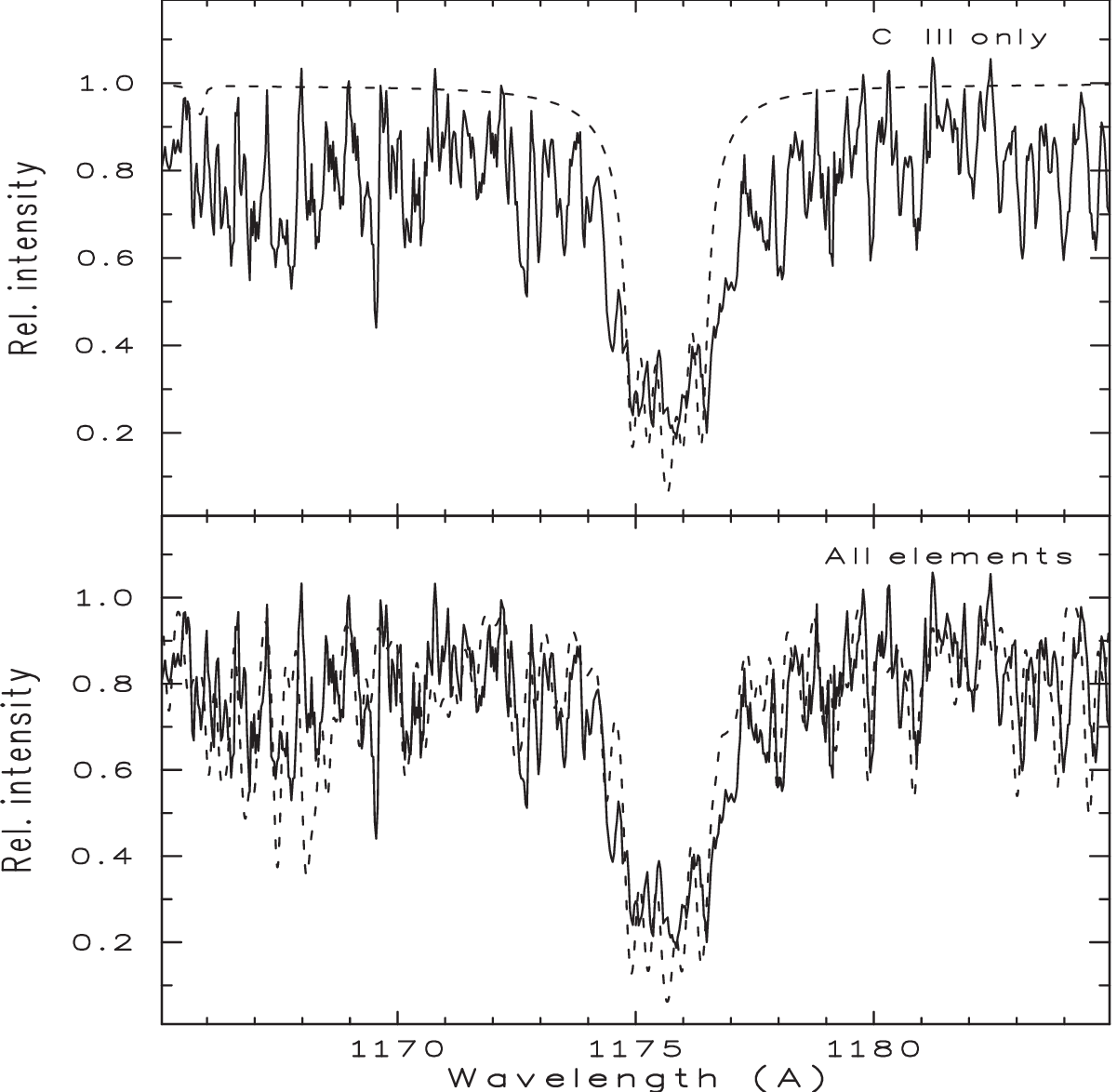}}
\resizebox{0.49\hsize}{!}{\includegraphics{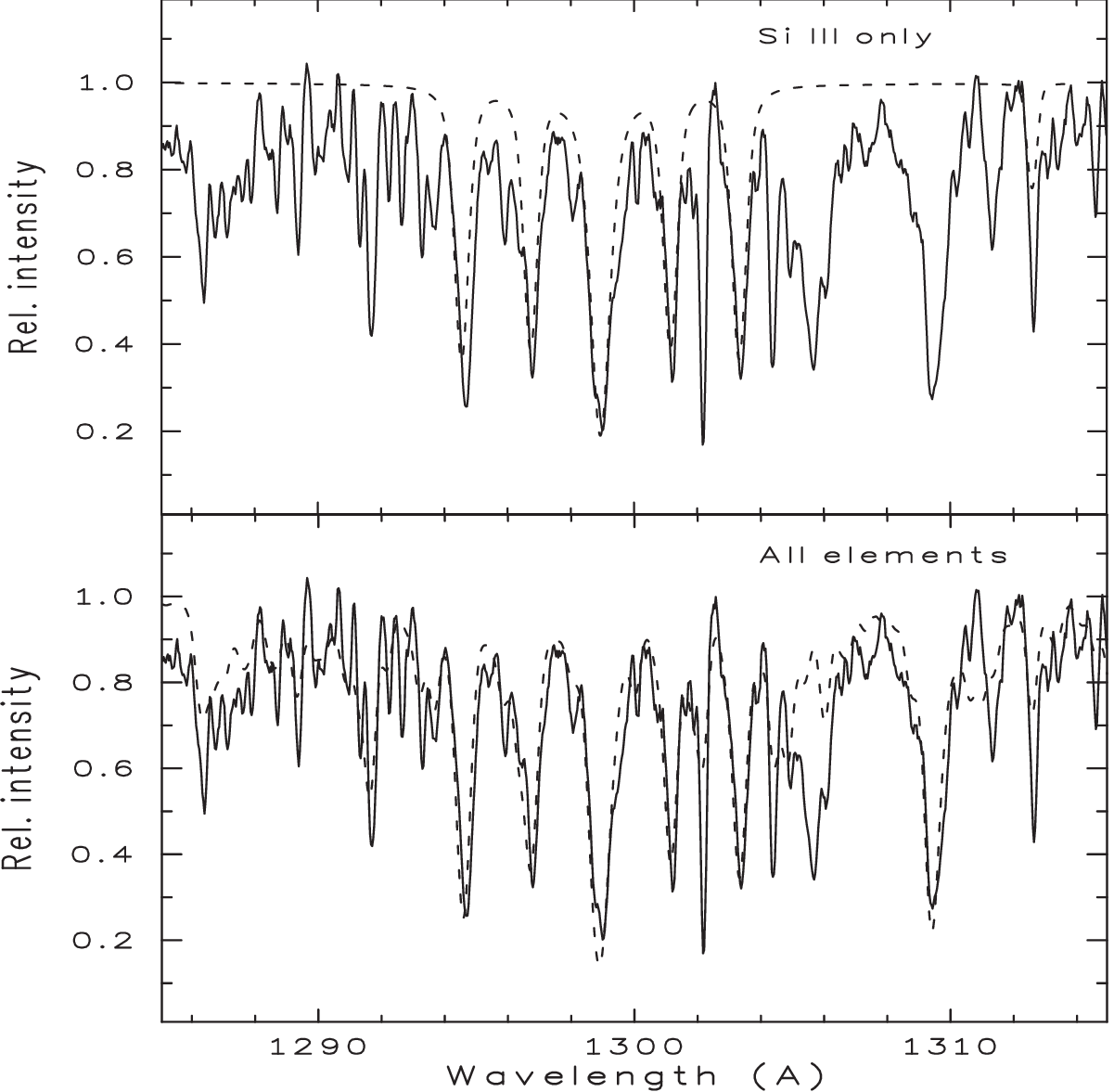}}
\caption{The average {\iue} high-dispersion spectrum of $a$~Cen (solid
 line) compared to synthetic spectra (dashed line). The synthetic spectrum labeled
 'All elements' was calculated with all elements. Left panel: The C~{\sc iii} only
 is synthetic spectrum showing only C~{\sc iii} lines at $\lambda$\,1175\,\AA.
 Right panel: The Si~{\sc iii} only is synthetic spectrum showing only Si~{\sc iii} lines at $\lambda$\,1300\,\AA.}
\label{ident}
\end{figure}

In order to minimize the uncertainties in the coefficients of the fitted curves,
the light curves were determined by averaging three nearest fluxes for a given $\lambda$:
\begin{equation}
 F(\lambda) = \frac{F(\lambda-\lambda_{step}) + F(\lambda) + F(\lambda+\lambda_{step})}
 {3},
\label{Mean_Flux}
\end{equation}
where $\lambda_{step}$ is equal 1.676\,\AA\ for SWP camera. As far as the errors in $F(\lambda)$ are concerned, we computed them by taking
into account the the errors in the fluxes as presented in INES $Catalog$,
according to the standard propagation theory of errors.

\sect{3 Identification of of C~{\bf III} and Si~{\bf III} depressions in
spectrum of $a$~Cen} \label{sect_3}

Figure~\ref{average} displays the average distribution of energy $(A_{0}(\lambda))$
of  $a$~Cen over the cycle of the variability in the spectral region from 1150 to 1950\,\AA. One can see from Fig.~\ref{average} that the deepest depressions of the flux are in the far-UV spectral region at $\lambda\lambda$\,1175.5, 1265, 1300 and 1335 \AA. It is well known that Si~{\sc ii} appears as the main absorber with the strong resonance lines at $\lambda$\,1260\,--\,64\,\AA\ \citep{Artru_Lanz1987} On the other hand, the depression at $\lambda$\,1335\,\AA\ appear from C~{\sc ii} resonance doublet at $\lambda$\,1334.5\,--\,1335.7\,\AA\ in the spectrum of $a$~Cen.

In order to identify which elements are responsible for depressions at $\lambda\lambda$\,1175.5 and 1300\,\AA\ the synthetic single-element spectra were calculated for C, Si and Fe with \citet{Piskunov_1992} program {\sc synth}. Mainly, the lines of these elements dominate in the far-UV spectral region of $a$~Cen.
Also, the synthetic spectra with all elements included were computed.
The information about spectral lines were taken from the Vienna
Atomic Line Database \citep[VALD-2,][]{Kupka_1999}.
Note that VALD-2 allows to compute the synthetic spectrum for stars
with abundances significantly different from the solar or scaled solar composition.
The adopted atmospheric parameters of $a$~Cen are the same as were used by
\citet{Bohlender2010}. The model computation was performed with $T_{\rm eff}$~=~19000~K, log~$g$~=~4.0 and $v_{\rm micro}$~=~2~km~${\rm s}^{\rm -1}$.
Experience showed that the spectral lines of twice ionized carbon and silicon mainly produce the deepest depressions at $\lambda\lambda$\,1175.5 and 1300 \AA, respectively.

\begin{table}
\begin{flushleft}
\centering
\caption{List of the spectral lines which produce depressions in the spectrum
 of $a$~Cen}
\smallskip
\label{Table_1}
\begin{tabular} {cccrr}
 \hline
Depression & Ion & Wavelength & E. P. & $\log{gf}$ \\ (\AA) & & (\AA) &(eV) \\
\hline
& C {\sc iii} & 1174.930 & 6.496 & -0.468 \\
& C {\sc iii} & 1175.260 & 6.493 & -0.565 \\
1175.5& C {\sc iii} & 1175.590 & 6.496 & -0.690 \\
& C {\sc iii} & 1175.710 & 6.503 & 0.009 \\
& C {\sc iii} & 1175.987 & 6.496 & -0.565 \\
& C {\sc iii} & 1176.370 & 6.503 & -0.468 \\
\hline
& Si {\sc iii} & 1294.545 & 6.553 & -0.037 \\
& Si {\sc iii} & 1296.726 & 6.537 & -0.127 \\
& Si {\sc iii} & 1298.892 & 6.553 & -0.257 \\
%& Si {\sc iii} & 1298.946 & 6.585 &  0.443 \\
1300.0 & Si {\sc iii} & 1301.149 & 6.553 & -0.127 \\
& Si {\sc iii} & 1303.323 & 6.585 & -0.037 \\
& Si {\sc ii} & 1304.370 & 0.000 & -0.423 \\
%& Si {\sc ii} & 1305.210 & 10.39 & +0.387 \\
%& Si {\sc ii} & 1305.592 & 6.859 & +0.710 \\
& Si {\sc ii} & 1309.276 & 0.036 & -0.123 \\
%& Si {\sc ii} & 1309.453 & 6.858 & +0.550 \\
%& Si {\sc ii} & 1309.725 & 6.859 & -0.590 \\
%& Si {\sc ii} & 1311.256 & 10.415 & +0.643 \\
& Si {\sc iii} & 1312.591 & 10.276 & -0.840 \\
\hline
\end{tabular}
\end{flushleft}
\end{table}

In order to confirm our hypothesis, the average high-dispersion spectrum was computed using {\iue} images SWP~14071, SWP~14080 and SWP~14088 which were obtained at phases 0.296, 0.393 and 0.509, respectively.
For all depressions the best agreement between the average {\iue} high-dispersion
spectrum of $a$~Cen and the synthetic spectra with all elements is reached
if the elements have solar scaled composition except for C and Si,
their abundances were reduced to $\log(N/N_{\rm total})$ of --4.0 and --5.0,
respectively. A comparison of the average {\iue} high-dispersion spectrum of $a$~Cen with full synthetic spectrum as well as those including only lines of C~{\sc iii} shows that this element is responsible for the depression of the flux at $\lambda$\,1175.5\,\AA\ (see Fig.~\ref{ident}). Practically, six C~{\sc iii} lines are responsible for depression of the flux at $\lambda$\,1175.5\,\AA. On the other hand, Si~{\sc iii} appears as the main absorber by the six strong resonance lines near $\lambda$\,1300\,\AA. Although, there is the some influence on this depression by Si~{\sc ii} resonance doublet at $\lambda$\,1304\,--\,1309\,\AA.
The list of these spectral lines are given in Table~\ref{Table_1}.

\sect{3 Results}

To measure the absorption in the cores of the depressions at $\lambda$\,1175.5 and 1300\,\AA\ we have formed the photometric indices $a_{1175}$ and $a_{1300}$, expressed in magnitudes:
\begin{eqnarray}
 a_{1175}={1 \over 2}(m_{1165}+m_{1185})-m_{1175}, \\
 a_{1300}={1 \over 2}(m_{1281}+m_{1318})-m_{1300},
\end{eqnarray}
where all filters are $\sim$5\,\AA\ wide for the $a_{1175}$ and the $a_{1300}$ indices.
These indices are analogous to the $a_{1400}$ index of \citet{Shore_Brown1987}.
The choice of the filter wavelengths was dictated by the low-dispersion {\it IUE} spectra of $a$~Cen, although the average high-dispersion spectrum indicate that
some filters can be affected by small spectral lines, as is illustrated by Fig.~\ref{ident}.

\begin{figure}[b]
%\vspace{-4.5cm}
\centering
\resizebox{0.47\hsize}{!}{\includegraphics{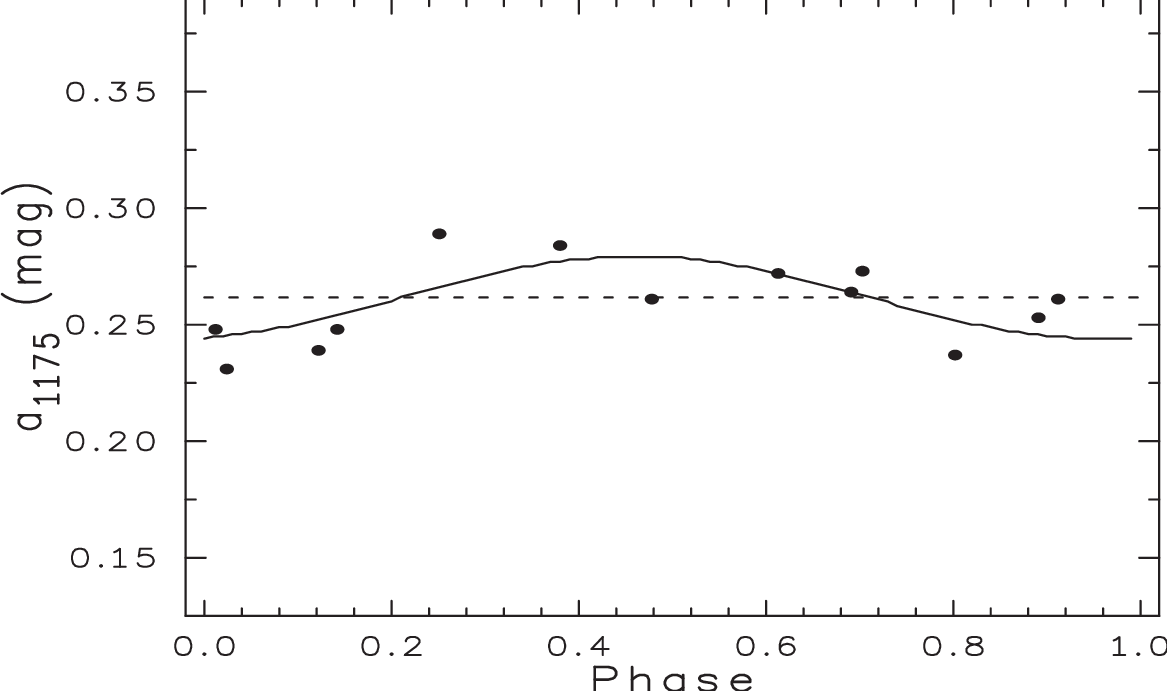}}
\resizebox{0.47\hsize}{!}{\includegraphics{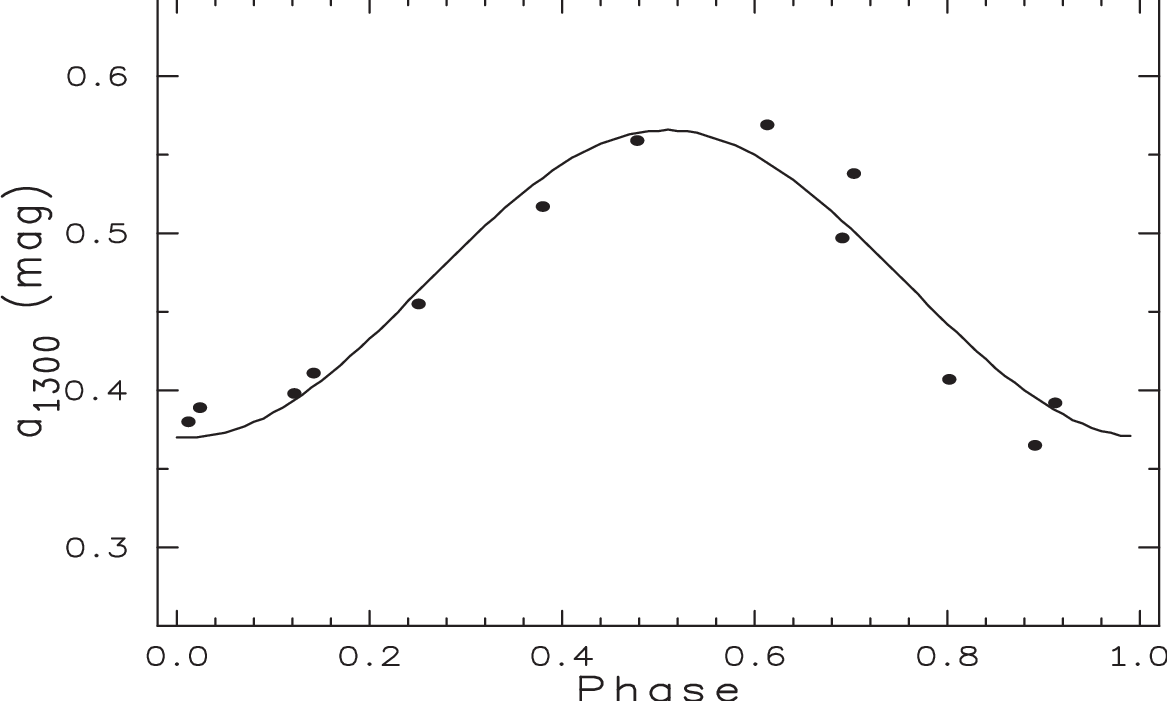}}
\caption{Left panel: Phase diagram of the $a_{1175}$ index.
 Right panel: Phase diagram of the $a_{1300}$ index.
 Solid lines are the least-squares fits. Dashed line indicate
 the average value of $a_{1175}$ index.}
\label{index}
\end{figure}

Figure~\ref{index} exhibits the variations of the measured absorption in the depressions at $\lambda\lambda$\,1175 and 1300\,\AA\ versus the rotational phase.
The solid lines represent least-squares fits by one-frequency cosine functions.
It should be noted that the phases were computed according to the new determinations of the ephemeris obtained by \citet{Sokolov2011}.
First of all, the least-square fit gives the semi-amplitude of the $a_{1175}$
index which is equal to 0.018~mag. But, the standard deviation of the residual scatter around the fitted curve is equal to 0.014~mag.
In other words, the photometric index $a_{1175}$ does not vary within errors of measurements, as is illustrated by Fig.~\ref{index}.
On the other hand, the photometric index $a_{1300}$ varies significantly.
Thus, the least-square fit gives the semi-amplitude of the $a_{1300}$ index which
is equal to 0.098~mag. Although, the standard deviation of the residual scatter around the fitted curve is equal only 0.023~mag.

For the first time, \citet{Norris_1971} has noted that the lines of N~{\sc ii}, Si~{\sc iii} and Fe~{\sc iii} vary at anti-phase with helium lines.
Most significant is the fact that while Si~{\sc iii} $\lambda$\,4552 varies by
50~m\AA, the line Si~{\sc ii} $\lambda$\,4130 does not vary at all.
Our result also indicates that Si~{\sc iii} lines centered on $\lambda$\,1300\,\AA\
in the ultraviolet spectral region vary at anti-phase with helium lines in the visual spectral region.
It should be noted that Si~{\sc iii} lines centered on $\lambda$\,1300\,\AA\ varies much more than Si~{\sc iii} line $\lambda$\,4552 in the visual spectral region.
In conclusion, an additional investigation is needed in order to do a final conclusion about the variability of $a$~Cen in the ultraviolet spectral region.

\end{document}